\documentclass[11pt]{article}

\usepackage[utf8]{inputenc}
\usepackage[T1]{fontenc}
\usepackage[margin=1in]{geometry}
\usepackage{graphicx}
\graphicspath{{Images/}{./}}
\usepackage{multirow}
\usepackage{amsmath,amssymb,amsfonts}
\usepackage{amsthm}
\usepackage{mathrsfs}
\usepackage{mathtools}
\usepackage{xcolor}
\usepackage{textcomp}
\usepackage{booktabs}
\usepackage{bm}
\usepackage{enumitem}
\usepackage{nicefrac}
\usepackage{authblk}

\usepackage[hidelinks]{hyperref}
\usepackage{orcidlink}

\DeclareSymbolFont{newfont}{OML}{cmm}{m}{it}
\DeclareMathSymbol{\Epsilon}{3}{newfont}{15}
\DeclareMathSymbol{\Varrho}{3}{newfont}{37}

\DeclarePairedDelimiterX\braket[2]{\langle}{\rangle}{#1\,\delimsize\vert\,\mathopen{}#2}

\title{Finite Element Simulation of Microwave Technologies for Powder-Based
Volumetric Additive Manufacturing Processes}

\author[1]{Brendan Mackey\,\orcidlink{0009-0002-6509-1196}}
\author[2,3]{Levi Smith\,\orcidlink{0000-0001-5996-5745}}
\author[1,3,4]{Bosco Yu\,\orcidlink{0000-0002-6786-3367}\thanks{Corresponding author: boscoyu@uvic.ca}}

\affil[1]{Department of Mechanical Engineering, University of Victoria, Engineering Office Wing, Victoria, BC V8P 3E6, Canada}
\affil[2]{Department of Electrical and Computer Engineering, University of Victoria, Engineering Office Wing, Victoria, BC V8P 5C2, Canada}
\affil[3]{Center for Advanced Materials and Related Technology (CAMTEC), University of Victoria, 3800 Finnerty Rd., Victoria, BC V8P 5C2, Canada}
\affil[4]{Institute for Integrated Energy Systems (IESVic), University of Victoria, 3800 Finnerty Rd., Victoria, BC V8P 5C2, Canada}

\date{}

\begin{document}

\maketitle

\begin{abstract}
Volumetric additive manufacturing (VAM) is a promising field of advanced manufacturing which reduces production times of conventional 3D printing while retaining application to complex geometries. Current VAM systems are limited to photo-polymerizing resins, which reduces the general applicability across industry. This work presents exploratory finite element simulations of microwave-based VAM systems compatible with ceramic, metal, or composite powder feedstock. The simulations explore three devices -- a linear resonator, a cross interferometer, and a boundary heater -- each with potential uses in manufacturing processes for one or multiple of the considered feedstock materials. We find that with precise control of input frequency, power, and waveguide transverse modes, thermal lattices can be engineered to selectively heat regions of the feedstock, allowing direct application to VAM. Resultant lattices have cellular dimensions on the order of 1 mm, and require input powers on the order of 100 kW. With advances in microwave technologies allowing for this precision, such devices could open pathways to VAM with materials currently outside the scope of optical methods.
\end{abstract}

\noindent\textbf{Keywords:} Volumetric additive manufacturing; microwave devices; finite element simulation; architected materials

\section{Introduction}\label{sec:intro}

Advances in manufacturing techniques have enabled the efficient production of precision components. In particular, additive manufacturing (AM) allows for reduced cost and lessened material waste for low-production parts and complex geometries \cite{zhou_additive_2024}, with applications across a myriad of industries including aerospace \cite{schiller_additive_2015, alami_additive_2023}, automotive \cite{alami_additive_2023, zhao_direct_2023, yang_application_2024}, construction \cite{bayat_effects_2025, paolini_additive_2019}, and healthcare \cite{amaya-rivas_future_2024, salmi_additive_2021}. Current AM technologies build structures by stacking two-dimensional layers of selectively extruded, photopolymerized, or fused feedstock \cite{zhou_additive_2024}. Such methods are convenient for prototyping, but often suffer from material limitations and lengthy production times reducing commercial viability \cite{kanishka_revolutionizing_2023}. Specifically, traditional polymer printing methods including stereolithography (SLA) \cite{huang_review_2020} and digital light processing (DLP) \cite{chaudhary_additive_2023} project 1D trajectories and 2D images respectively onto an imaging plane to solidify a layer of resin, with repeated layers producing a three dimensional part \cite{sampson_multimaterial_2021}. These approaches are inherently slow due to the small amount of resin illuminated and the inter-layer mechanical movements \cite{chaudhary_additive_2023, orth_automatic_2025}.

\begin{figure}[!htbp]
    \centering
    \includegraphics[width=0.75\textwidth]{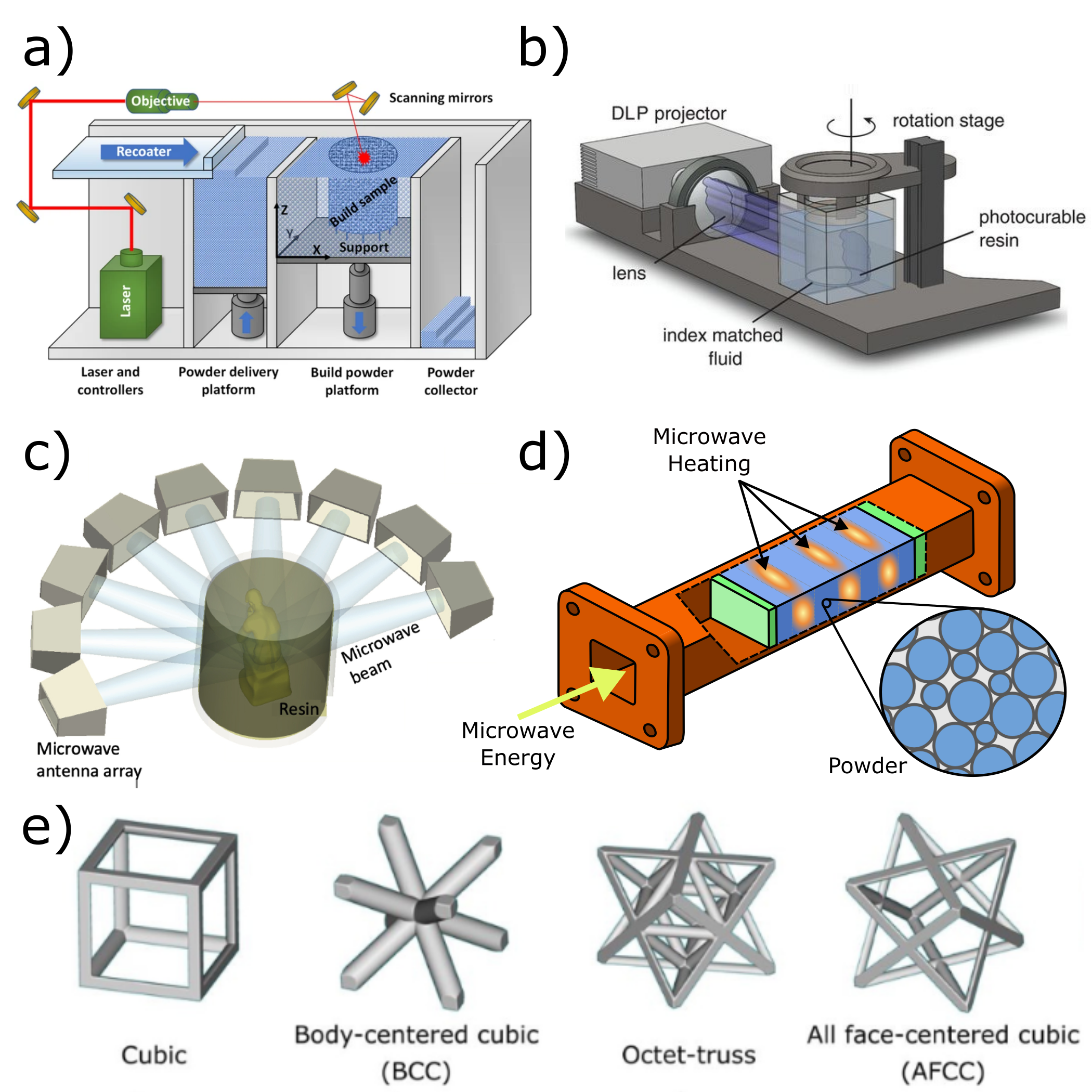}
        \caption{Examples of current and proposed AM systems. a) A graphic of laser PBF for metal powders \cite{bulgarevich_crystal_2023}. b) A schematic of a VAM system using photosensitive resins and optical frequencies. The DLP projector uses variable diffraction and interference from paraxial sources to produce time-dependent images along a single axis, and the rotation stage allows extension of a two-dimensional image plane to three-dimensional shapes. Figure from ``Volumetric additive manufacturing via tomographic reconstruction'' \cite{kelly_volumetric_2019}. Adapted with permission from AAAS. c) The microwave VAM system proposed by Mukherjee et al. \cite{mukherjee_towards_2024}, equivalent to the optical system in b). With a sufficient number of antennas, the antenna array provides complete control over the electric field intensity and can achieve VAM without a rotation stage. Figure from ``Towards microwave volumetric additive manufacturing: Generation of a computational multi-physics model for localized curing'' \cite{mukherjee_towards_2024}. Adapted with permission from Elsevier. d) A graphic of the proposed microwave VAM system. A waveguide (orange) containing a volume of powder (blue) held in place by dielectric walls (green), has microwave energy injected into the system. The natural electromagnetic modes of the device, dependent on input transverse modes, selectively melt regions of the powder where the field intensity is the highest. e) Selection of architected 3D lattice unit cells. This work focuses on lattices with unit cells around 1 mm in dimension. Figure from ``Architected cellular materials: A review on their mechanical properties towards fatigue-tolerant design and fabrication'' \cite{benedetti_architected_2021}. Adapted with permission from Elsevier.}
    \label{fig:intro}
\end{figure}

To address these issues, methods of expanding material compatibility and decreasing production times have been independently developed \cite{zhou_additive_2024}. Printing of metal components, an important frontier in advancing AM capabilities, has been achieved through powder bed fusion (PBF), illustrated in Fig.~\ref{fig:intro}a), binder jetting, and directed energy deposition (DED) \cite{patel_metal_2024, cooke_metal_2020, armstrong_overview_2022}. Binder jetting is a moderately efficient process, reaching build rates of up to 1000 L/h in industrial settings \cite{erhard_binder_2026, castroecosta_review_2017}, but struggles with precision due to shrinkage in the sintering process and deformation from process defects \cite{yu_research_2025, wang_investigation_2017}. Meanwhile, PBF (0.1 L/h) and DED (0.5 L/h) produce high-precision parts slowly \cite{dejene_current_2023, popov_powder_2021, avrampos_review_2022}. Beyond metals, composites using embedded nanomaterials have been explored to modify the material properties of conventional AM feedstocks \cite{rahman_review_2024}, bioprinting can extend AM capabilities to organic materials \cite{yilmaz_bioprinting_2021}, and select AM processes can be used with a limited range of functional ceramics\cite{olhero_conventional_2022}. Such approaches are in preliminary stages, however, and achievable material properties, manufacturing precision, and production efficiencies are not certain to surpass those of pure metals, ceramics, and glasses incompatible with these techniques \cite{mukalay_systematic_2025, sing_direct_2017}.

An alternative to vastly improve production efficiency is volumetric additive manufacturing (VAM) where whole volumes of feedstock are solidified simultaneously \cite{kelly_volumetric_2019, whyte_volumetric_2024, sampson_multimaterial_2021, orth_automatic_2025, bernal_road_2025}. VAM systems utilize interference and diffraction of optical lasers to form intensity patterns in photopolymerizing resin \cite{bernal_road_2025, bhattacharya_high_2021}. An example system is provided in Fig.~\ref{fig:intro}b), where a rotating stage is used to allow single-axis imaging \cite{kelly_volumetric_2019}. This process is significantly faster than two-dimensional printing processes as it illuminates entire volumes of resin and avoids the need for support material \cite{orth_automatic_2025}. However, this approach is limited to transparent, photosensitive feedstocks and optical scattering prevents use of filler-loaded or nanomaterial-loaded feedstocks \cite{bernal_road_2025, mukherjee_towards_2024}. These limitations can be partially avoided using hydrogel infusion on printed parts \cite{ji_volumetric_2025}, but infusion lengthens the required production times and reduces the achievable precision. Using microwave frequencies in addition to optical frequencies improves upon the compatible materials and processes \cite{pozar_microwave_2012}; microwaves can propagate in materials opaque to optical frequencies and experience little scattering from microscale particles \cite{kwak_microwave_2015, tsang_scattering_2000}. 
Recent work has explored the use of microwave antenna for VAM with filler-loaded composites \cite{mukherjee_towards_2024}, as shown in Fig.~\ref{fig:intro}c), successfully demonstrating applicability in engineering electrical, mechanical, and other chemical properties of printed components. These composite materials show great promise, but still require a resin matrix, and thus still limit the range and purity of materials compatible with VAM \cite{mukherjee_towards_2024}.

To our knowledge, no study has explored microwave VAM in resin-free powder feedstock. As a proof of concept, we perform a numerical feasibility study of microwave devices, extending this work to powder-based feedstock using ohmic and dipolar heating. A simple version of such a system is shown in Fig.~\ref{fig:intro}d), where a microwave waveguide device allows excitation of deterministic modes, so that the intensity pattern inside the powder is given by the device geometry and the choice of input mode and frequency. 

The natural electromagnetic modes of many devices produce periodic lattices \cite{jackson_classical_2009}, which have direct applications in the manufacturing of architected materials. Architected materials are porous materials formed by interconnected networks of solid struts, nodes, and voids \cite{ashby_mechanical_1983}. These materials can produce extreme stiffness to density ratios \cite{deshpande_effective_2001, schaedler_ultralight_2011}, high strength to density ratios \cite{bauer_high-strength_2014}, and provide excellent thermal insulation\cite{zhou_effective_2024} leading to direct applications in aerospace \cite{khan_systematic_2024, wang_evaluating_2020}, advanced automotive \cite{tian_research_2025}, and biomedical industries \cite{tian_research_2025, distefano_titanium_2023, kovacs_comparative_2023}. They are especially attractive in additive manufacturing because 1) their complex architectures are difficult to fabricate using traditional subtractive methods, 2) they enable topology optimization and offer lightweight solutions for structural applications such as aerospace load-bearing panels \cite{khan_systematic_2024, wang_evaluating_2020, akula_metallic_2026} and biomedical implants \cite{distefano_titanium_2023, kovacs_comparative_2023}, and 3) they simultaneously reduce the material usage, cost, and embodied energy of the manufactured part \cite{zhou_additive_2024}. One subclass of architected materials are 3D lattices; these are of particular interest to the AM community due to their natural compatibility with AM processes \cite{metallurgy_and_materials_society_of_additive_2025, korkmaz_development_2022}. A selection of such lattices is shown in Fig. \ref{fig:intro} e) for geometric reference.  In this work, we integrate the design of resonators, interferometers, and waveguide systems to produce fast, energy efficient methods of producing lattices in ceramic-based materials and shells in metals.

\section{Scope of Study}\label{sec:scope}
This study is a proof-of-concept and feasibility analysis of microwave VAM in powder feedstocks. Using finite element methods (FEM), we identify three device architectures with potential for use in AM: 1) a linear resonator leveraging internal field enhancement and standing wave modes, 2) a cross interferometer using interference fringes to achieve sub-wavelength resolutions, and 3) a boundary heater to extend material compatibility to metal powders for specific geometries. Each addresses a distinct technical challenge. The linear resonator presents a possible method for reducing power requirements. Cavity field enhancement raises the internal intensity for a given source power, at the cost of restricting the printable patterns to the discrete set of resonant modes. The cross interferometer gives up that enhancement in exchange for pattern freedom, since arbitrary combinations of input modes and phases can be injected, and it delivers finer feature spacing. The boundary heater addresses feedstock compatibility, abandoning volumetric three-dimensional patterns so that metal powders, which cannot be heated internally, remain accessible as hollow shell geometries.

The work uses FEM to integrate analysis of electromagnetic resonator structures with heating and heat transfer. We narrow the scope of the study through constraints on the electromagnetic frequencies, feedstock materials, and simulated lattice structures considered. The requirements on microwave power, input transverse mode, and physical device imperfections are left free, and the identified requirements are benchmarked against existing systems.

We focus on super high frequency (SHF) and millimeter wave frequencies (between 8 GHz and 40 GHz). This band was selected for its millimetre scale wavelengths, while avoiding scattering off micron dimensioned powders \cite{hecht_optics_2017}. This balance allows us to apply an effective medium approximation critical to defining the electromagnetic and thermal properties of the powder feedstock, while still permitting sub-millimeter resolution in the phase structures.

The feedstock materials are limited to insulating ceramics, metals, and composite materials formed from mixtures of metal and ceramic powders. Applying this constraint narrows the design space, creating a simplified methodology for defining the electromagnetic fields. Notably, semiconductors are excluded from this study; they are known to have highly temperature sensitive properties \cite{samara_temperature_1983} with limited empirical data available for high temperatures.

The target thermal lattice structures are regularly spaced nodes of high temperature -- near the melting temperatures of the selected feedstock to model sintering processes. These structures arise naturally from the modal profiles of many microwave devices. This simple example is used as a proof of concept, as combinations of nodal patterns allow the formation of near-arbitrary lattices.

Additionally, due to high power requirements, all investigated systems fully contain modes within conducting layers. This addresses safety issues from microwave absorption and Paschen breakdown in uncontrolled atmosphere \cite{chen_introduction_2018, kuffel_high_2009} limiting the results to systems with direct experimental potential. 

\section{Methods}\label{sec:theory}
\begin{figure}[!htbp]
    \centering
    \includegraphics[width=\textwidth]{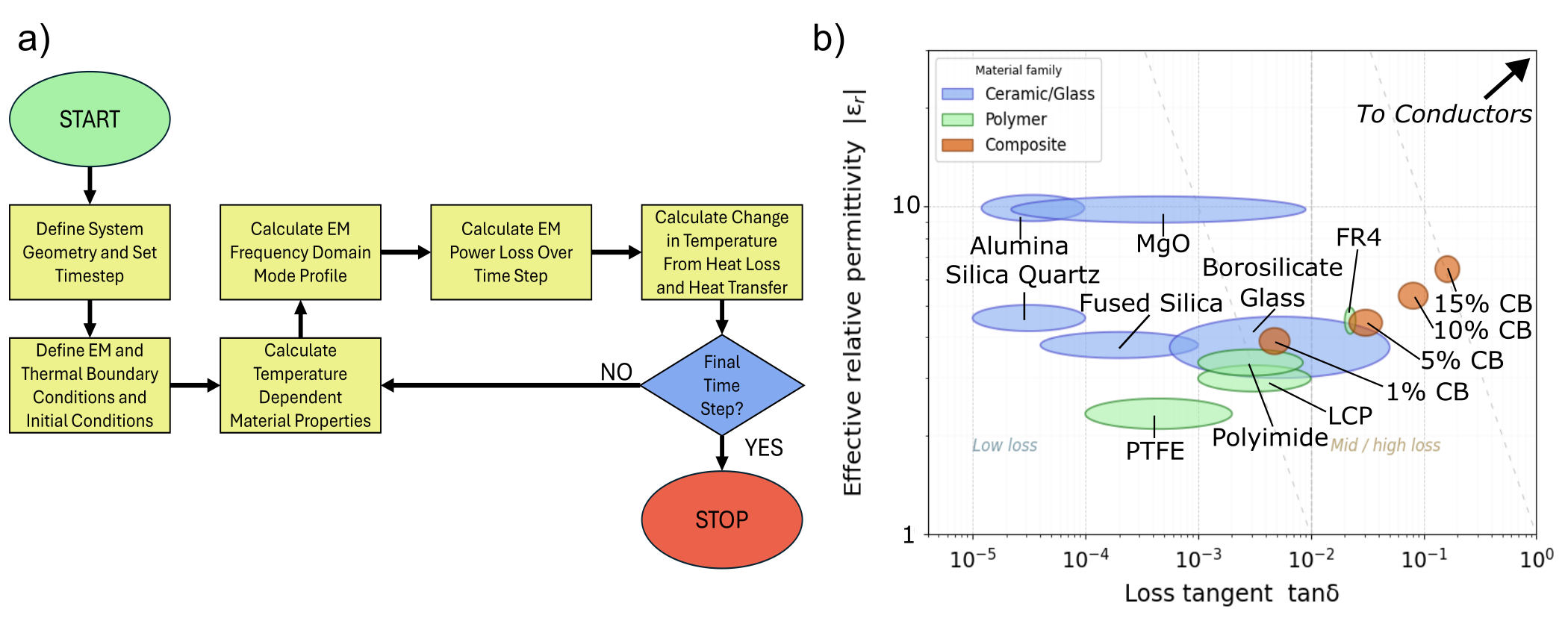}
   \caption{a) A flowchart of frequency-transient simulation process. The iterative loop from \textit{Calculate Temperature Dependent Material Properties} to \textit{Final Time Step} is automated to be completed inside a single COMSOL simulation, without any user initialization. b) Material properties chart of loss tangent against effective permittivity at 10 GHz frequencies and room temperature. The materials are grouped into ceramics/glasses, polymers, and composites. Data for the ceramics (alumina\cite{mbango_coaxial_2025, penn_effect_1997}, silica quartz\cite{zou_novel_2023}, fused silica\cite{shearer_glassceramic_2025, mbango_coaxial_2025}, borosilicate glass \cite{shearer_glassceramic_2025, humphrys_effect_1941}, and magnesium oxide [MgO] \cite{mazierska_precise_2005}) and polymers (polytetrafluoroethylene [PTFE] \cite{shearer_glassceramic_2025, zou_novel_2023}, polyimide \cite{liaw_advanced_2012}, liquid crystal polymers [LCP] \cite{zhou_flexible_2022}, and FR4\cite{sofin_r_g_measurement_2005}) come from experimental data. Notably, MgO is an outlier where the data was taken at 80K ($-193^\circ C$). The composites are calculated using the Bruggemen Effective mean approximation for a mixture of fused silica and carbon black [CB] with the given percentage representing the volume fraction of carbon black. Metals/conductors are ignored as the relevant properties are both extremely high ($|\varepsilon_r|, \mathrm{tan}\delta \gg 1$), and the materials are best represented by conductivity, not permittivity.}
    \label{fig:Flowchart}
\end{figure}

Finite element simulations are conducted in COMSOL Multiphysics \cite{comsol} using the RF and heat transfer modules. Due to the difference in timescales between the electromagnetic mode formation and heat transfer dynamics, a compound frequency-transient simulation is used to evaluate the electromagnetic modal properties in the frequency domain, which are in turn used as heat sources for a heat transfer analysis in the time domain. Theory of relevant electromagnetic devices and microwave heating is provided in Supplemental Information Section 1 and Section 2 respectively. The interplay of these phenomena requires an iterative loop, where the electromagnetic and thermal state at each time step influences the next states. A flowchart describing the general simulation algorithm is provided in Fig.~\ref{fig:Flowchart}a). 

In addition, an effective medium approximation is applied to improve simulation efficiency. Powder-based materials are physically non-homogeneous at the scale of a powder grain dimension. As such, accurate representation requires a mesh much smaller than the powder dimension, which drastically increases the computational power required. The alternative method, described in Supplemental Information Section 3, finds that powders with dimensions much less than the microwave wavelength can be treated as a single bulk medium allowing for optimized simulations while preserving physical accuracy. This method applies both to homogeneous powders and mixtures of multiple materials, also providing a pathway to engineering material properties of the feedstock, illustrated by the loss tangent against permittivity chart in Fig.~\ref{fig:Flowchart}b) displaying a selection of dielectric materials considered in simulations. Because material properties are temperature-dependent (Supplemental Information Section 4), the heat source evolves in time. 

Our method allows complete characterization of the electromagnetic heating, within the numerical accuracy dictated by the spatial mesh size and the timestep interval. Unless otherwise specified, all simulations use free tetrahedral elements with a maximum cell dimension of $\lambda/10$ for the wavelength  $\lambda$ in each material region, and apply impedance boundary conditions at all waveguide walls to capture ohmic losses. 

Allowable timescales are identified for two distinct elements of the simulation: the discretized time step and the total simulation time. The temperature profiles change with each time step and thus the electric field profiles will jump between discrete profiles dependent on the jump in permittivity $\varepsilon_r(\mathbf{r})$. To avoid non-physicality, a sufficiently small time-step (such that $\varepsilon_r$ changes by $\leq 0.1\%$ over a time-step) is maintained for smooth mode evolution and deterministic thermal patterns. Convergence testing is run to ensure that the temperature profiles are independent of spatial discretization and timestep, even for values below the analytic threshold, which is discussed further in Supplemental Information Section 5.

Similarly, contrast between the hot antinodes and the cold nodes survives only while the thermal diffusion length $\sqrt{4\kappa t}$ remains well below the antinode spacing \cite{carslaw_conduction_1959}. Here, $\kappa$ is the thermal diffusivity and $t$ is the total simulation time. For fused silica, $\kappa \approx 7\times10^{-7}$ m$^2$s$^{-1}$ and is nearly constant from room temperature to the glass transition \cite{hofmeister_effects_2012}, so the diffusion length reaches 0.84 mm over the 250 ms heating time of Sec.~\ref{subsec:Fabry-Perot}, an order of magnitude below the antinode spacing. This bounds on the usable timescales for total heating and, thus, the required powers to reach specific temperatures while retaining thermal lattice structures. The same maximum temperature ($T_{max}$) could in principle be reached with a tenth of the power over ten times the duration, but the resulting part would be a uniformly heated block rather than a lattice.

\section{Results}\label{sec:results}
The specific designs and performance specifications of each device are discussed in the sections below. Importantly, each design identified required high microwave power, ranging from 200kW to 1MW, to prevent lattice distortion from heat-transfer dynamics discussed above.

\subsection{The Linear Resonator}\label{subsec:Fabry-Perot}
The linear resonator design is a Fabry-Perot resonator built into a waveguide. This device produces standing waves that form 3D lattices in the electric field intensity within low-to-moderate absorption dielectric powders, using the combination of transverse modes induced by the waveguide boundaries and longitudinal mode induced by the Fabry-Perot boundaries. This could offer a direct pathway to printing basic lattice structures in ceramic and composite powders, with structures limited to superpositions of resonant mode sets.

\begin{figure}[!htbp]
    \centering
    \includegraphics[width=0.75\textwidth]{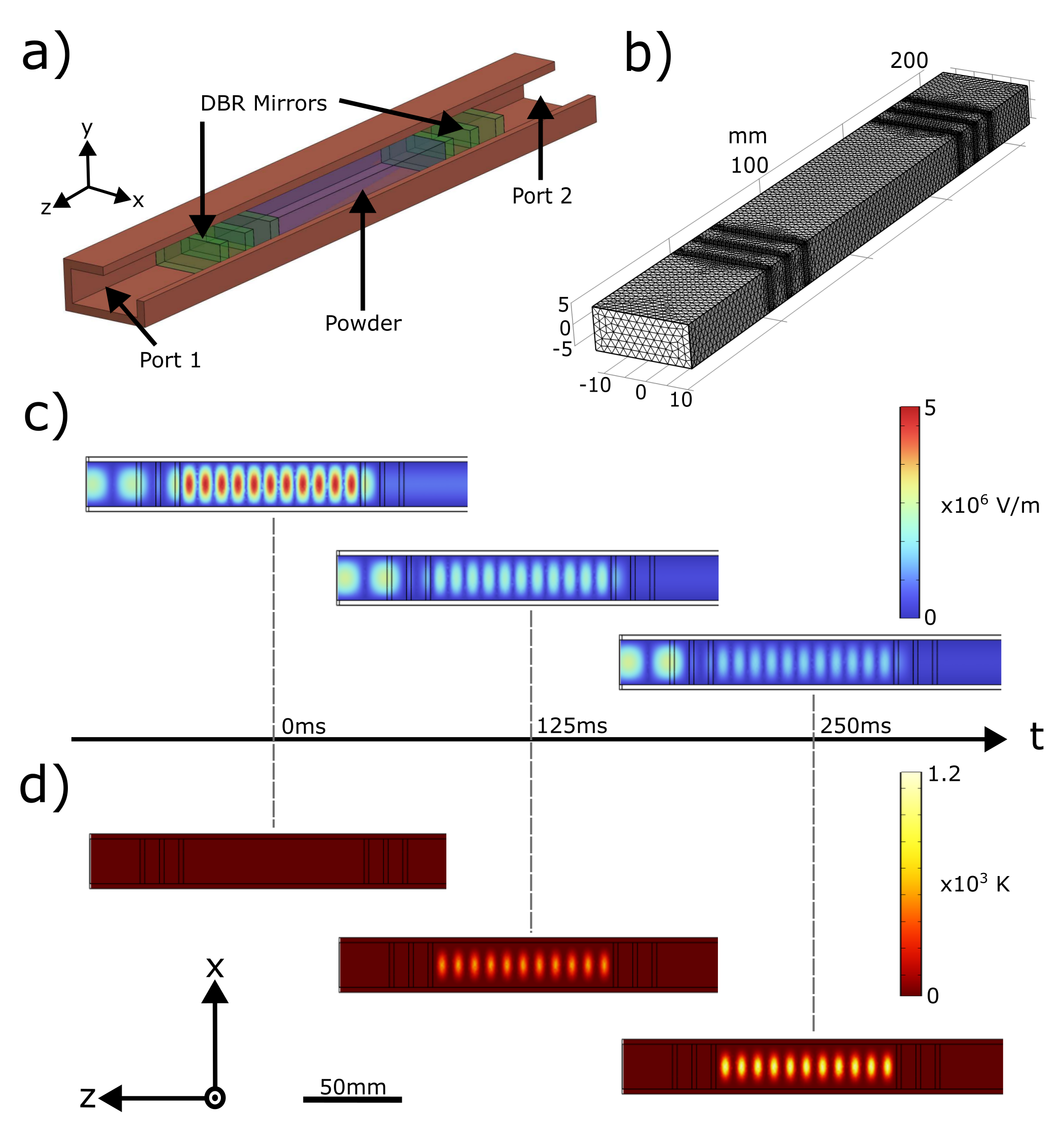}
   \caption{Schematic and simulation of the linear resonator model with pure fused silica feedstock. a) A diagram of the system showing the resonator components. Port 1 is an active input port and port 2 is a receiver port. b) The tetrahedral mesh profile generated for RF and thermal simulation of a). The difference in mesh sizing for each region originates from the varied $\varepsilon'$ values between layers. c) A two-dimensional slice at $y = 0$ of the calculated electric field profile as a function of time. An input power of 250 kW was used. d) Resultant temperature distribution from the electromagnetic heating in the same plane c). The heating was calculated in increments of 0.1 ms, to update the electric field profile at each step, over 250 ms total.}
    \label{fig:Fabry-Perot}
\end{figure}

The schematic, shown in Fig.~\ref{fig:Fabry-Perot}a), separates the device into regions for the distributed Bragg reflector (DBR) mirrors and the cavity containing the powder. Each DBR mirror consists of layers of high-purity alumina ($\varepsilon_r = 9.7$, $\mathrm{tan}\delta = 10^{-4}$ with the center stopband at 10 GHz \cite{woode_measurement_1994}) separated by air gaps.
The powder is fused silica, due to the low thermal dependence of permittivity ($\alpha \approx 5\mathrm{x}10^{-5}$) and the larger loss tangent $\mathrm{tan}\delta \approx 10^{-3}$, an order of magnitude above both silica quartz and alumina \cite{zhu_investigation_2009, lide_physical_2005}. The relative permittivity is given as 3.8 at room temperature \cite{lide_physical_2005}. The powder-filled cavity region is 97.3 mm long. Including the field penetration into the DBR mirrors, the effective optical length of the resonator is $6\lambda_g \approx 99.9$ mm at the 9.825 GHz resonance, where $\lambda_g$ is the guided wavelength in the effective medium; the difference corresponds to approximately 1.3 mm of penetration at each mirror. 

In this model, the temperature dependence of the silica permittivity follows a linear extrapolation (described in Supplemental Information Section 4), applied over the full range from room temperature to the melting temperature of silica, $T_m = 1986 K$ ($1713^\circ C$) \cite{lide_physical_2005}, with the loss tangent and thermal properties taken from the activation-energy and interpolated-table models of the same section. Since measured permittivity data for silica approaching $T_m$ are scarce, this extrapolation is the principal source of quantitative uncertainty in the results, which motivates the sensitivity study of Sec.~\ref{subsec:Cross}.

Turning to the finite element parameters, the resulting inhomogeneous mesh is shown in Fig.~\ref{fig:Fabry-Perot}b); its resolution differs significantly between the alumina layers and the airgaps, and the time step interval was set to $0.1$ms. Microwaves, in the fundamental mode $TE10$, were injected through an active port at one end of the waveguide and collected at both the injection port and a receiver port at the other end of the waveguide.

The time-domain results, represented in Fig.~\ref{fig:Fabry-Perot}c) for electric field and Fig.~\ref{fig:Fabry-Perot}d) for the temperature, show predictable heating at the antinodes of the resonator standing waves, with derived spacing of $8.33\,\mathrm{mm} = \lambda_g/2$, where $\lambda_g$ is the guided wavelength of the powder-filled waveguide.  

The internal field enhancement reduces the power requirements to achieve heating without significant lattice distortion from heat transfer. The three-period DBR system shown in Fig.~\ref{fig:Fabry-Perot} requires 250 kW of input power to achieve $T_{max} = 1215$K ($942^\circ C$) in $250$ ms, while an identical system with the DBR mirrors removed only heated to $T_{max} = 393K$ ($120^\circ C$) in the same time interval. Here $T_{max}$ is an outcome rather than a design target. It is the temperature at which the competition between field enhancement and the temperature-based field filtering described below saturates.

\begin{figure}[!htbp]
    \centering
    \includegraphics[width=0.75\textwidth]{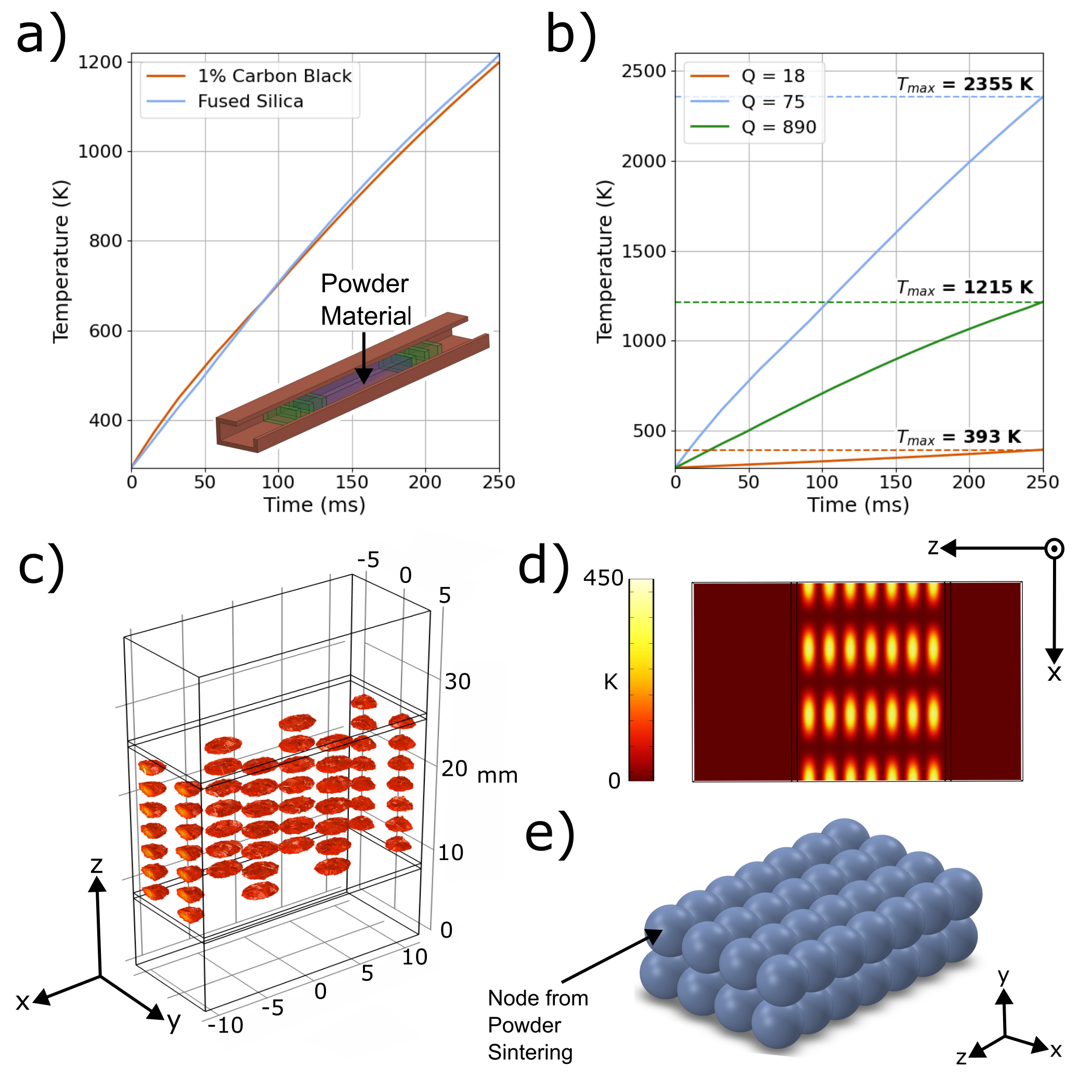}
   \caption{Investigations of maximum achievable internal temperature over 250 ms and 3D temperature lattice formation. a) Effect of the effective-medium loss tangent on the heating curve, comparing pure fused silica ($\mathrm{tan}\delta = 0.001$) with a 1\% carbon-black composite ($\mathrm{tan}\delta = 0.02$). b) Effect of resonator Q-factor on the heating curve; the Q-factors $Q = 18, 75,$ and $890$ correspond to zero, one, and three alumina/air periods per mirror respectively. c) Shows the 3D lattice inside a single-layer resonator, and identifies the minimum longitudinal and transverse spacings to be 2.1 mm and 5.1 mm. d) Depicts a x-z slice of temperature intersecting the point of maximum temperature in the volume ($T_{max} = 443K$ [$170^\circ C$]), detailing the longitudinal lattice. e) Depicts a rendering of the architected material output from this device. Using pulsed heating sequences with applied heat transfer, the thermal nodes broaden to an overlapping profile, with struts defined by the transverse mode profile, and sinter a cubic (or rectangular) lattice structure.}
    \label{fig:QvTemp}
\end{figure}

However, this time dependent electric field profile illustrates a key behaviour of this device: temperature-based field filtering. The permittivity of the effective medium monotonically increases with temperature, altering the phase gained over the cavity length. Gradually, the cavity phase is pushed out of the linewidth resonance condition for the resonator, and thus the internal field decays. This competition between internal field enhancement and decay time naturally limits the maximum achievable temperatures $T_{max}$ by the sensitivity of the medium's permittivity, a potential concern with this device. Adjusting the simulation to include a 1\% carbon-black additive in the feedstock, yielding an effective $\varepsilon'_{eff} = 3.75$ and $\mathrm{tan}\delta_{eff} = 0.02$ at room temperature, validated this behavior. The heating achieved a similar $T_{max} = 1198K$ ($925^\circ C$), but initially heated at a faster rate. The comparison of heating times is shown in Fig.~\ref{fig:QvTemp}a), demonstrating the similarity of temperature curves despite enhanced absorption. The increased heating rate, however, reduces the required power to avoid heat-transfer distortions, and drops the power requirements to 210 kW for heating to the same $T_{max}$ over 250 ms.

The system in Fig.~\ref{fig:Fabry-Perot} is extended to varied cavity Q-factors in Fig.~\ref{fig:QvTemp}b), where the temperature profile with time is given as a function of Q-factor for Fabry-Perot resonators. Notably, the $T_{max}$ is highest for the $Q = 75$ resonator. This exemplifies the tradeoff between internal field enhancement and linewidth, identifying weak resonators as optimal for such systems. However, this value exceeds the melting temperature of silica and is non-physical under the linearized permittivity model. The practical implication is that moderate-Q resonators offer the best balance between field enhancement and linewidth before temperature-filtering effects dominate. As temperatures far exceeding $T_m$ are undesirable for printing, the $T_{max} \approx T_m$ provides a practical limitation in selecting the ideal resonator strength or printing time.

Additionally, the temperature filtering effect also presents strong advantages, as it constrains the internal fields to a nearly constant profile while heating. This improves the lattice quality with higher-Q resonators.

\begin{figure}[!htbp]
    \centering
    \includegraphics[width=0.75\textwidth]{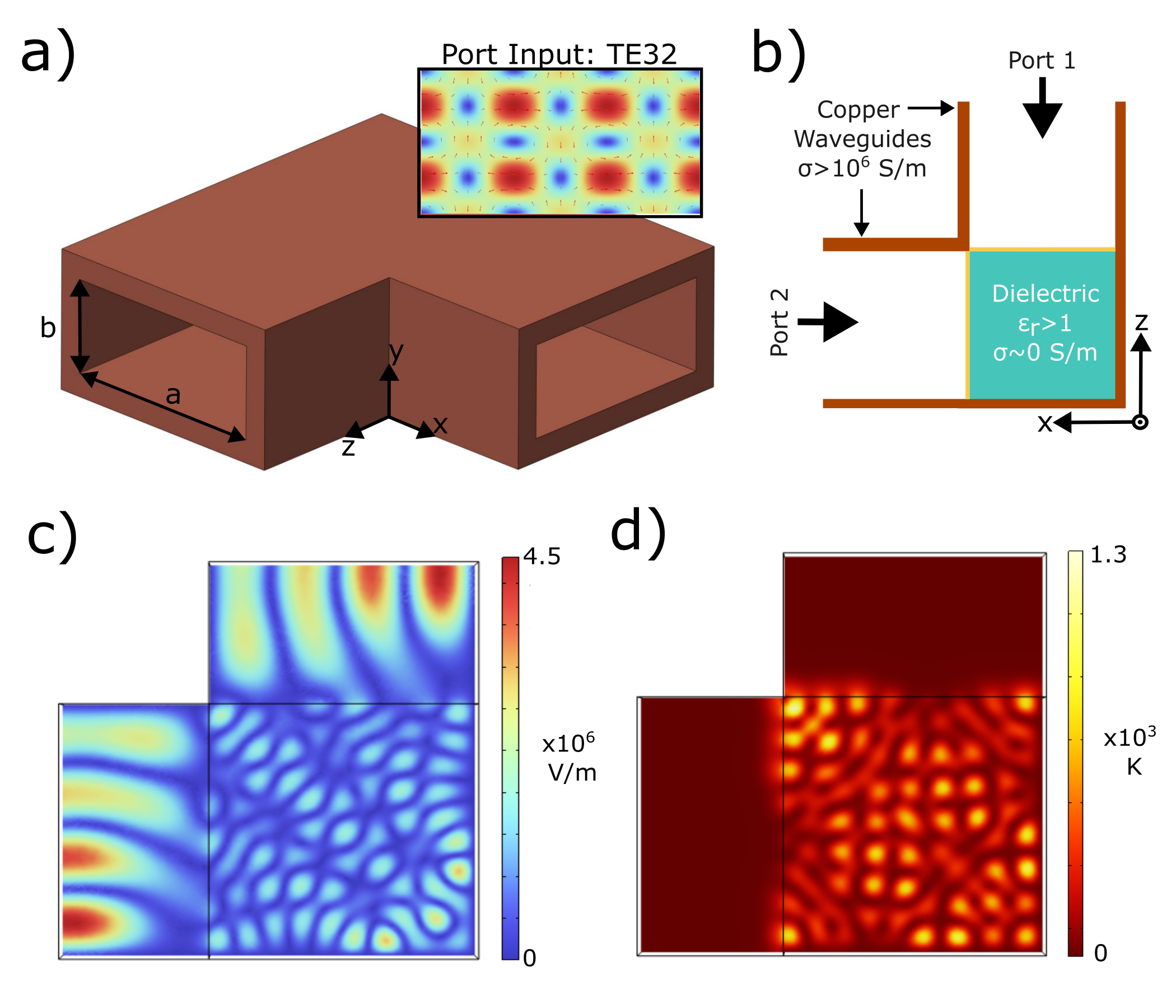}
   \caption{Schematic and simulation of the cross interferometer model for a $TE32$ input mode. a) A diagram of the system, with the inset providing the input mode. The dimensions a and b are matched to a WR-90 (a = 0.9in x b = 0.4in) waveguide input. b) A cross sectional detail view of the schematic in a). The dielectric region is the absorptive powder, with the thin yellow walls representing the alumina. Both port 1 and port 2 are active input ports. c) A two-dimensional slice at $y = 0$ of the calculated electric field profile at 100 ms. d) Resultant temperature distribution from the electromagnetic heating in the same plane c). The heating was calculated in increments of 0.1 ms with the electromagnetic mode profile recomputed at each increment.}
    \label{fig:cross}
\end{figure}

Use of the $TE10$ mode produces a one-dimensional, longitudinal thermal lattice pattern. Extension to higher order modes, such as the $TE32$ mode, produces three-dimensional thermal lattices in the internal field. A single reflector layer (Q = 75) resonator tuned to resonance for the $TE32$ mode at 36 GHz produces the 3D thermal lattice in Fig.~\ref{fig:QvTemp}c), with an x-z slice highlighting the complete thermal pattern shown in Fig.~\ref{fig:QvTemp}d). The increase in frequency to 36 GHz, accounts for the higher cutoff frequency for the $TE32$ mode \cite{jackson_classical_2009}. This opens pathways to deterministic lattice printing with linear resonators. Using pulsed heating sequences could produce regular lattices of sintered material, such as the simple cubic lattice shown for this specific system in Fig.~\ref{fig:QvTemp}e). The natural applications for these outputs include ceramic lattice cores for thermal insulation, high-temperature filters or catalyst supports where a periodic open-cell ceramic architecture is the functional requirement, and lightweighting bulk materials of any dielectric. 

Additionally, the resonance condition and the difference in transverse mode propagation constants \cite{jackson_classical_2009, pozar_microwave_2012} can be used to filter out all non-desirable transverse modes, effectively preventing lattice distortions from multimode excitation. A complement to this mode filtering, a power-weighted superposition of resonant modes (at other frequencies or transverse mode orders) could be input to further engineer the thermal lattices. Practical implementation of resonators with linewidths narrow enough to isolate specific modes would require a high degree of tunability of the cavity length. Thus, the feasibility of these devices would be limited by the temperature sensitivity of the powder materials and the engineered tuning length of the cavity.

\begin{figure}[!htbp]
    \centering
    \includegraphics[width=\textwidth]{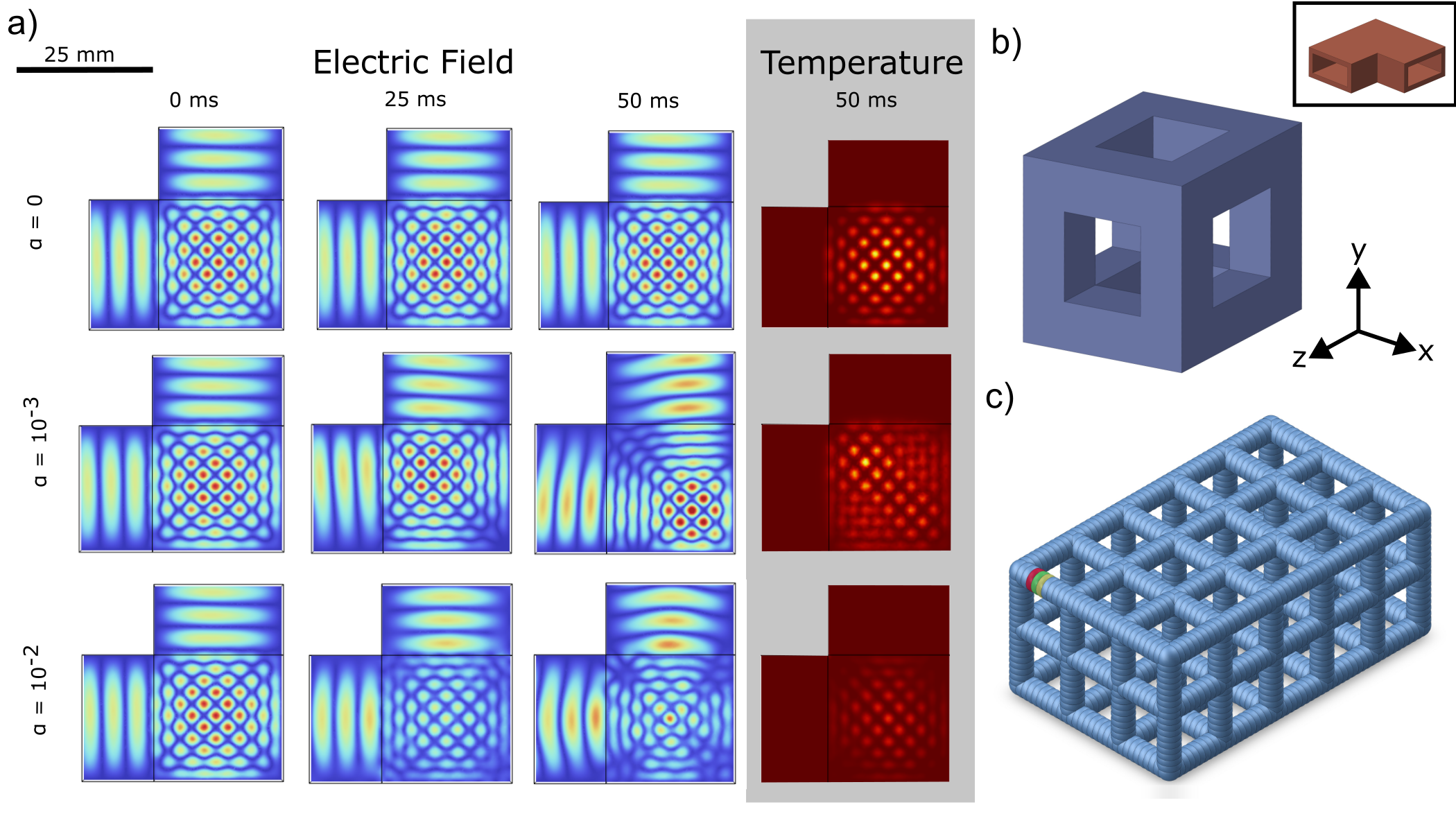}
   \caption{A sensitivity study and ideal device output using frequency sweeps for the cross interferometer (shown in inset). a) A study of the sensitivity of the heating profile to linear permittivity temperature coefficient $\alpha$. Each simulation used the $TE10$ mode over 50 ms to model the electric field stability and generated thermal profile. b) A render of an ideal simple cubic strut-based unit cell output from this device. c) A render of realistic device output formed by sintering over independent frequency sweeps at each port. The highlighted sections (red, green, yellow) outline sintered regions from independent thermal nodes formed at different input frequencies.}
    \label{fig:tempsens}
\end{figure}

\subsection{The Cross Interferometer}\label{subsec:Cross}
The cross interferometer design is an L-joint connecting two WR-90 waveguides, with a basin of powder held in the overlapping regions by two thin alumina walls. This creates three-dimensional lattices by interfering inputs from two perpendicular waveguides, analogous to a four-source antenna array (with the input and back reflection all contributing to the electric field profile). This design allows for arbitrary inputs, in comparison to the set of resonant modes of the linear resonator, and thus can produce more controllable field profiles while losing possible advantage from internal field enhancement and mode stabilization. 

The schematic, shown in Fig.~\ref{fig:cross}a), and cross sectional view, shown in Fig.~\ref{fig:cross}b), depict the system geometry. Importantly, the model includes a small fillet at both right angle corners to prevent field singularities with increased mesh resolution. In this model, the dielectric powder is the same fused silica as in the linear resonator model, with the properties adjusted for 36 GHz frequencies \cite{naftaly_terahertz_2021}. Microwaves, in $TE32$ mode, were injected into both Port 1 and Port 2 (shown in Fig.~\ref{fig:cross}b) to form three-dimensional interference patterns in the powder region. The simulation was conducted over $100$ms in $0.1$ms intervals, the duration being chosen to satisfy the heating timescale requirement set out in Sec.~\ref{sec:theory} while resolving the drift of the interference pattern.

The electromagnetic field in the powder region varied significantly with temperature, leading to a continuous change of the interference pattern and, thus, heated regions. Such modal instability created strong deviation between the electric field profile, shown in Fig.~\ref{fig:cross}c), and temperature profile, shown in Fig.~\ref{fig:cross}d), at the final timestep. 

Despite this, the temperature profile still demonstrated regions of lattice formation with $1.86$ mm spacing, again leveraging the $TE32$ mode profile to improve lattice resolution, and $T_{max} = 1232 K$ ($959^\circ C$) for two $100$ kW port inputs and $100$ ms heating time. The formation of the lattice was numerically stable and could be repeatably predicted, despite the non-constant field profiles. The practical problem this raises is that the mapping from the input field -- its frequency, mode content, and port phases -- to the resulting thermal pattern is deterministic but not intuitive, so a user cannot read off the required excitation from a desired part geometry. Any future user of a microwave VAM system must therefore solve this inverse problem numerically before printing. As such, methods of inverse design, such as the Gerchberg–Saxton algorithm \cite{gerchberg_holography_1972, fiddy_legacies_2013}, could be extended to time-dependent material properties for the calculation of mode-stabilizing pulse sequences.

The above results demonstrate the ability to use numerical simulations to predict the thermal profile. The applicability of these predictions to physical systems, however, depends largely on the stochasticity of the system response to change in material properties. In dielectrics, this requires quantifying the sensitivity of the mode shape to $\varepsilon_r (T)$. The spacing between interference fringes is, predictably, highly dependent on $\varepsilon_r$ at room temperature. However, the geometric shape does not change significantly and the $\varepsilon_r$ of a material can be experimentally measured for calibration \cite{krupka_microwave_2021, wang_fast_2021}. More important is quantification of the effect of mode disturbance due to position dependent permittivity within the powder region. For this, simulations of the cross interferometer were completed for four non-physical linear proportionality constants $\alpha$: 0$K^{-1}$, $10^{-4}K^{-1}$, $10^{-3}K^{-1}$, and $10^{-2}K^{-1}$. The results, for $TE10$ input modes, are given in Fig.~\ref{fig:tempsens} a), which shows the three cases $\alpha = 0$, $10^{-3}$, and $10^{-2}$ K$^{-1}$. Here, it is observed that the electric fields vary greatly over time for $\alpha = 10^{-3},\,10^{-2}$ but the final temperature distributions all take similar geometric profiles. The field stochasticity contributed to a broadening of the temperature profile, reducing $T_{max}$ and increasing the background heating. This suggests the performance of experimental systems may be lower than predictions, but small variance in material properties would not invalidate the physical relevance of the simulations.

The primary advantage of these systems over the previous system is tunability of the internal profile, due to the electromagnetic field depending on interference of input modes rather than a device geometry alone. As such, the device could produce a range of different lattice structures from sintered material. For example, a strut-based simple cubic lattice (unit cell geometry shown in Fig.~\ref{fig:tempsens}b)), could be produced by scanning the input frequencies at each port with the other port fixed at a common frequency defining the cell size. Fig.~\ref{fig:tempsens} c) shows the resulting output geometry formed by repeated sintering the thermal nodes through the frequency scan. However, the geometry in Fig.~\ref{fig:tempsens}c) does assume similar permittivities between high-temperature powder and the post sintered component. These structures could have similar applications in ceramic heat exchangers, catalyst supports,  and filter media, but they also extend to capabilities to a wider range of architected dielectric materials. Here, the selected lattice geometries could produce fine control over mechanical, thermal, or electrical bulk properties. 

\begin{figure}[!htbp]
    \centering
    \includegraphics[width=0.8\textwidth]{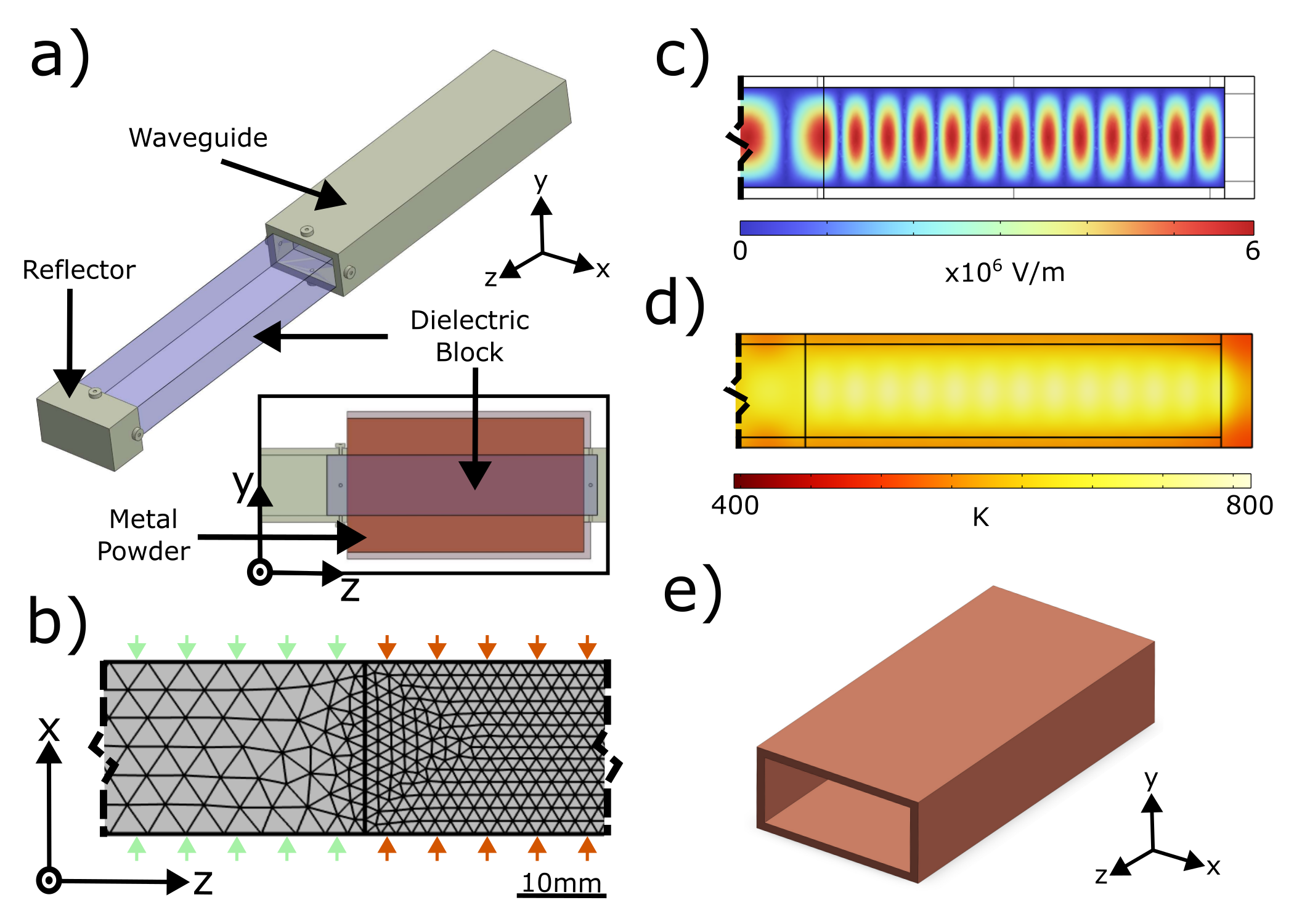}
   \caption{Schematic and simulation of the external boundary heater model. a) A diagram of the system showing the reflector system. The inset shows how metal powder enforces a boundary condition around the dielectric block, effectively active as a waveguide. The length of the dielectric block was optimized for maximum field enhancement in the silica at 10 GHz. b) The tetrahedral mesh profile generated for RF and thermal simulation of a). The green arrows show the impedance boundary conditions for waveguide material (tungsten) and the red arrows for powder material (copper).  c) A two-dimensional slice at $y = 0$ and $t = 1500$ ms of the electric field profile. An input power of 1MW was used. d) Resultant temperature distribution from the electromagnetic heating along the boundary (powder) wall. The heating was calculated in increments of 1 ms. e) Render of the copper shell produced by this device with TE20 input mode and a short heating time.}
    \label{fig:extprint}
\end{figure}

\subsection{The External Boundary Heater}\label{subsec:ExtPrint}
Both of the previous designs limit the compatible powder properties to dielectrics. Extending usage to pure metal powders requires a shift of approach from internal volumetric heating to boundary heating. This motivates the introduction of this external boundary heater, where metal powder acts as a waveguide boundary condition and is heated via induced eddy currents. This process provides an alternative path to powder-based printing with pure metal powders but limits the geometries to hollow shells. As such, the boundary printer is more akin to casting, where the geometry of the device acts as a fundamental limitation on the shell shape. 

The schematic, shown in Fig.~\ref{fig:extprint}a), depicts the geometry and powder interaction. As shown in the inset, the device is placed within a powder basin such that all exposed faces of the dielectric block contact the metal powder. Here, input power is retro-reflected off the metal plate at the end, forming a standing wave from a single round trip rather than from a resonant build-up.

In the electromagnetic simulation, we model the metal powder as a copper impedance boundary and the solid waveguides as a tungsten impedance boundary. Despite the lower conductivity, the tungsten boundaries heat slower than the copper boundaries due to tungsten's larger volumetric heat capacity \cite{white_heat_1984}. Similarly, contact points between the hot regions of copper or silica and the tungsten will not induce melting of the waveguides, as tungsten's $T_m = 3,695 K$ ($3,422^\circ C$) far exceeds copper's $T_m = 1,358 K$ ($1,085^\circ C$). The boundary conditions and the resulting mesh are shown in Fig.~\ref{fig:extprint}b). The dielectric block is a high-purity silica quartz, using low loss ($\mathrm{tan}\delta = 10^{-4}$) to concentrate energy loss at the boundaries and high temperature stability ($\alpha = 10^{-4} K^{-1}$) to maintain a stable electric mode profile. This simulation was coupled to the thermal simulation by surface heat sources linked to the impedance boundaries. Importantly, the thermal simulation included the entirety of all waveguides and extended into the powder basin to accurately model heat transfer dynamics. The simulation used 1MW of input power in the $TE10$ mode over a heating time of $1500$ms in $1$ms intervals.

The electric field mode, shown in Fig.~\ref{fig:extprint}c) for $t = 1500$ms has notable stability over the heating duration. This stability originates from the insignificant change in skin depth between the maximum and minimum conductivity, effectively maintaining constant boundary conditions while enhancing energy loss. The mode stability produces a constant heating profile along the inside of the system wall, shown in Fig.~\ref{fig:extprint}d). This leads to a $T_{max}= 789 K$ ($516^\circ C$). However, the high thermal conductivity of the metals increases the rates of heat transfer and thus requires extremely high input powers to prevent dissipation from the thermal lattice sites.

Alternative to the node based mesh structures, superpositions of several transverse modes can yield near uniform temperature distributions at the powder-enforced boundaries. Angled surfaces can also produce more uniform heating at lower input power requirements due to mode mixing and impedance matching. Simulated results of the angled surfaces are discussed further in Supplemental Information Section 6.

The stable mode and high thermal conductivity lead to native production of shells of sintered metal powder around the outer surface of the waveguide. For the rectangular geometry discussed, the sintered part is a thin-walled rectangular lattice, shown in Fig.~\ref{fig:extprint}e), if the heating time is short and a thin-walled solid tube if the heating time is long. The wall thicknesses are set by the input power and heating time. This creates a process more akin to casting than to 3D printing, but could provide an efficient method of manufacturing thin-walled enclosures, ducting, and hollow waveguides.

\section{Conclusion}\label{sec:con}
This work presents exploratory simulations of three complementary microwave-based VAM architectures: 1) a linear Fabry-Perot resonator, 2) a cross interferometer, and 3) an external boundary heater. Here we demonstrate the potential of each to extend VAM processes to feedstocks incompatible with optical VAM. The linear resonator produces deterministic thermal lattices in dielectric powders through the interplay of transverse and longitudinal modes. This system is limited by temperature-based field filtering, which reduces the achievable heating rates but also prevents lattice distortion from geometrically changing electric fields. The cross interferometer extends this capability to complex interference patterns, but reduces the modal stability. The external boundary heater allows compatibility with pure metal powders through ohmic heating at waveguide boundaries, but reduces the potential geometries to hollow shells. In every case the part is formed by driving the antinodes above the sintering threshold while the surrounding powder stays loose and is removed afterwards. The simulations stop at the onset of densification: as porosity is eliminated and melting begins, the permittivity and loss tangent of the feedstock change in a way that the linearized material model does not capture, and high-temperature data for feedstock is scarce. Measuring this change in permittivity and loss tangent through densification is the single most valuable material input for future experimental implementation.

Together, these systems establish promising pathways for designing microwave VAM apparatus across the full range of ceramic, metallic, and composite feedstocks. The natural mode stability of the resonator and retro-reflector designs will allow for direct application of inverse design processes to produce complex lattice geometries. In contrast, the cross interferometer provides the most control over the internal mode profile at the cost of mode stability. However, this instability is deterministic and thus these simulations could be used to produce pulse sequences of varying frequency and mode to stabilize the geometric mode profile. Additionally, more input ports could be added to the cross interferometer design to provide even greater control of lattice profiles and higher resolution than demonstrated in this work.

Furthermore, this work also reveals the limiting factors for experimental implementation. All three systems required high input powers (200 kW to 1 MW) to overcome heat transfer away from the thermal lattice sites in fused silica. While MW microwave powers are achievable, with the Canon E3718 Klystron as an example for GHz frequencies, they are expensive and large, with the E3718 having a length of 3.5 m and mass 1250 kg. These power requirements could be avoided by identifying and characterizing the temperature dependence of materials with higher loss tangents, such as the composite powders discussed in this work. Additionally, current limitations on selective transverse mode excitation also present technical challenges for implementation; these may be addressed by the selective mode excitation methods currently being explored in academia and industry. With these improvements, these designs may present a feasible and efficient method to produce architected materials.

Here, we demonstrated deterministic and non-chaotic heating profiles across several devices. Further work exploring inverse design for these devices, including introducing time-dependence for material properties, could lead to productive advances in simulation and design of practically achievable hardware. Alternatively, work in characterization of the electromagnetic properties of various materials at high temperatures would enhance the predictive capability of the simulations. The reliance on approximate temperature-dependent material models introduces uncertainty in quantitative predictions, though the sensitivity study for the cross interferometer indicates that the geometric form of the thermal lattice is robust to moderate ($10^{-3} -10^{-2}$) variations in the permittivity coefficient $\alpha$. This study demonstrates that microwave VAM holds real promise as a route to volumetric printing of ceramics, composites, and metals that lie beyond the reach of optical methods. With higher loss tangent composite feedstocks and the mode-selective microwave sources now emerging, a first experimental demonstration of powder-based VAM may be a realistic goal.

\section*{CRediT authorship contribution statement}
\textbf{Brendan Mackey:} Conceptualization, Methodology, Software, Formal analysis, Investigation, Visualization, Writing -- original draft,  Writing -- review \& editing.
\textbf{Levi Smith:} Conceptualization, Supervision, Funding acquisition, Resources, Project administration, Writing -- review \& editing.
\textbf{Bosco Yu:} Conceptualization, Supervision, Funding acquisition, Resources, Project administration, Writing -- review \& editing.

\section*{Declaration of competing interest}
The authors declare that they have no known competing financial interests or personal relationships that could have appeared to influence the work reported in this paper.

\section*{Acknowledgements}
Financial support from the Natural Sciences and Engineering Research Council of Canada Discovery Grants (B. Yu: RGPIN-2023-04456 and L. Smith: RGPIN-2022-03277) is gratefully acknowledged.

\section*{Data availability}
The simulation datasets from the study are available from the corresponding author on reasonable request.

\section*{Supplementary material}
Supplementary information accompanies this article.

\clearpage
\appendix
\renewcommand{\thesection}{S\arabic{section}}
\renewcommand{\theequation}{S\arabic{equation}}
\renewcommand{\thefigure}{S\arabic{figure}}
\renewcommand{\thetable}{S\arabic{table}}
\setcounter{section}{0}
\setcounter{equation}{0}
\setcounter{figure}{0}
\setcounter{table}{0}
\renewcommand{\theHsection}{S\arabic{section}}
\renewcommand{\theHequation}{S\arabic{equation}}
\renewcommand{\theHfigure}{S\arabic{figure}}
\renewcommand{\theHtable}{S\arabic{table}}

\begin{center}
{\large\bfseries Supplementary Information: Finite Element Simulation of Microwave Technologies for Powder-Based Volumetric Additive Manufacturing Processes}
\end{center}

This document provides supplementary information supporting the main text,``Finite Element Simulation of Microwave Technologies for Powder-Based Volumetric Additive Manufacturing Processes''. This document includes theoretical background, numerical convergence validation, and additional simulation results.

\section{Microwave Waveguides and Resonators}\label{subsec:waveguide}
Waveguides and resonators are key elements of microwave systems, which define the geometric modes of the electromagnetic fields. The modal properties of microwave propagation can be described completely by the electric field profile satisfying \cite{bondeson_computational_2005}

\begin{equation}
    \label{eq:emsolver}
    \nabla \times (\mu_r^{-1} \nabla \times\mathbf{E}) - \frac{\omega^2}{c^2}\left( \varepsilon_r - \frac{i\sigma}{\omega \varepsilon_0}\right)\mathbf{E} = 0
\end{equation}

\noindent at all points. Here, $\mu_r$ is the relative permeability, $\varepsilon_r$ and $\varepsilon_0$ are the relative and vacuum permittivity respectively, $\sigma$ is the material conductivity, $c$ is the speed of light in vacuum, and $\omega$ is the microwave frequency. In arbitrary structures, the modal properties can be found by applying equation \eqref{eq:emsolver} through an FEM.

\begin{figure}[!htbp]
    \centering
    \includegraphics[width=0.75\textwidth]{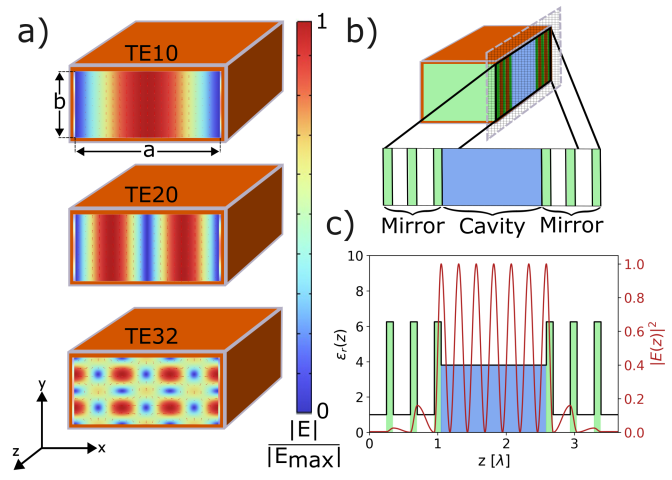}
    \caption{ Diagrams of microwave waveguides and resonators showing electromagnetic mode profiles. a) Rectangular WR-90 waveguides with inset normalized transverse field profiles for TE10, TE20, and TE32 modes. The dimensions are $a = 0.9$ in and $b = 0.4$ in by supplier specification. b) Schematic of a Fabry-Perot Cavity built inside a waveguide. The detail view provided is taken along the center axis of the waveguide. The blue region represents the cavity dielectric material, the green represents the reflector material, and white represents air. c) Normalized standing wave mode for the schematic in b). The mode profile is represented by the field intensity, given by the red curve, and the relative permittivity details the physical structure by the black curve.}
    \label{fig:EMDevice}
\end{figure}

Waveguides are a useful structure with analytic solutions to their natural mode sets \cite{pozar_microwave_2012}. Commonly, they are rectangular or cylindrical metal tubes which provide boundary conditions in two transverse directions (x and y) through the metal walls, while leaving the longitudinal (z) direction unconstrained. This separation of transverse and longitudinal conditions reduces to a paraxial system, where a set of transverse modes satisfy equation \eqref{eq:emsolver} for the waveguide boundary conditions \cite{jackson_classical_2009} independent of the longitudinal mode. This produces a set of transverse fields, shown in Fig.~\ref{fig:EMDevice}a), which can propagate through the waveguide structure. For rectangular waveguides, these are separated into transverse electric (TE) and transverse magnetic (TM), with TEml denoting the TE mode of orders m,l. Any combination of these modes can, in principle, be selectively excited to form deterministic intensity patterns within the waveguides and engineer the interactions with resonators or other devices. Modern technologies struggle with selective excitation of higher-order waveguide modes ($m,l \geq 2$), but research is ongoing to improve this functionality with prospects in phase-selective devices \cite{lauterio-cruz_higher_2023, ruskuc_excitation_2018} and microwave analogs to spatial light modulation using controllable metasurfaces \cite{wu_intelligent_2019, cui_coding_2014}. Unless otherwise specified, all waveguides in this work are copper WR-90 (0.4in x 0.9in cross section) to use a commercially available product compatible with the selected frequency range of study.

Resonators, particularly low-finesse Fabry-Perot resonators \cite{jackson_classical_2009}, form standing waves by applying weak boundary conditions on the longitudinal mode. A Fabry-Perot resonator, shown in Fig.~\ref{fig:EMDevice}b), consists of two partially reflective, inwards facing mirrors with a gap (cavity) between them \cite{hecht_optics_2017, pozar_microwave_2012}. The one simple mirror commonly used in resonators is a Distributed Bragg Reflector (DBR) mirror \cite{saleh_fundamentals_2001}. These consist of layers of alternating dispersive materials with $\lambda/4$ thickness, for $\lambda$ being the wavelength in the medium, where interference effects between the layers enhance reflection. Aperture mirrors are also commonly integrated with waveguide technologies \cite{pozar_microwave_2012}, but they are not discussed further in this work. Using resonators in waveguides permits enhancement of internal fields. In cavities, which satisfy the round-trip phase condition $\phi_{rt} = \beta L = 2\pi n$ for propagation constant $\beta$, cavity length $L$, and integer $n$ for all points on a wavefront, the counter-propagating traveling waves form internal standing waves, and thus increase all light-matter interactions at antinodes of the modal patterns. The amount of internal field enhancement can be quantified by the Q-factor $Q = \nicefrac{\omega}{\Delta \omega}$ \cite{hecht_optics_2017}, a metric of energy stored in the resonator mode. This relationship highlights the largest limitation for implementing resonator systems; the inverse proportionality with the linewidth $\Delta \omega$ dictates the spread of frequencies that can form quasi-resonant cavity modes. In simple terms, the larger the internal field enhancement, the smaller the range of frequencies the resonator will accept, reducing tunability and increasing the importance of thermal/mechanical stability. The Q-factor is proportional to the reflectivity of the resonator mirrors, and so varying the number of DBR-layers -- where more layers lead to higher reflection -- allows direct control of the Q-factor. The one-dimensional calculated intensity pattern for the cavity in Fig.~\ref{fig:EMDevice}b) is given in Fig.~\ref{fig:EMDevice}c). Recent work has demonstrated effective methods of engineering the standing wave modes for optical frequencies \cite{ossiander_metasurface-stabilized_2023}, and similar methods in the microwave regime provide control over the intracavity intensity patterns.

\section{Microwave Heating and Current Induction}\label{subsec:heat}
The majority of microwave heating occurs due to two distinct mechanisms: dielectric heating and ohmic heating. Dielectric heating occurs when the oscillating electric field drives molecular polarization out of phase \cite{metaxas_industrial_2008, sun_review_2016}. These losses cause dipolar rotation and drive the movement of charges, causing various forms of resistive heating in polar molecules, disordered lattices, and powders. The heat generation per unit volume can be given by \cite{sun_review_2016}

\begin{equation}
    \label{eq:dielectric_heat}
\left(\frac{dP}{dV}\right)_{di} = \frac{\omega}{2}\varepsilon_0\varepsilon'' |\mathbf{E}|^2
\end{equation}
where $\varepsilon''$ is the imaginary component of permittivity $\varepsilon_r = \varepsilon' -j\varepsilon''$. Generally, $\varepsilon''$ is small for non-polar, ordered lattices with values as low as $\mathrm{tan}\delta \approx 2.7\mathrm{x}10^{-5}$ for alumina \cite{alford_sintered_1996}. This loss can become significant for polar molecules, such as epoxy with $\mathrm{tan}\delta = 0.52$ \cite{calisir_designing_2025}, and can increase due to Maxwell-Wagner polarization \cite{metaxas_industrial_2008} in the presence of defects and material interfaces, leading to a sizable difference between silica quartz ($\mathrm{tan}\delta \approx 2\mathrm{x}10^{-4}$) and fused silica ($\mathrm{tan}\delta \approx 10^{-3}$) \cite{zhu_investigation_2009, lide_physical_2005}. Importantly, Maxwell-Wagner polarization suggests that the absorption of bulk dielectric might be an underestimation of the absorption of a powdered dielectric due to the concentration of material surfaces.

In ohmic heating, in contrast, the oscillating field drives currents through conductive materials and energy is dissipated by resistive heating \cite{sun_review_2016}. In this regime,

\begin{equation}
    \label{eq:ohmic_heat}
\left(\frac{dP}{dV}\right)_{ohm} = \frac{\sigma_{eff}}{2} |\mathbf{E}|^2
\end{equation}
where $\sigma_{eff}$ is the effective conductivity of a conductor or composite mixture. This heating is particularly relevant to metals, where the skin depth $\delta = \sqrt{\nicefrac{2}{\mu \sigma \omega}}$ determines the heating volume \cite{jackson_classical_2009}. Both dipolar heating and ohmic heating are natively captured by the loss tangent \cite{pozar_microwave_2012}

\begin{equation}
    \label{eq:loss_tangent}
\mathrm{tan}\delta = \frac{\varepsilon''}{\varepsilon'} + \frac{\sigma}{\omega \varepsilon_0 \varepsilon'}
\end{equation}
which fully quantifies electric field absorption as $\varepsilon_r = \varepsilon'(1 -j\mathrm{tan}\delta)$.

With heating power and location determined by the field profile and material properties, the total heating profile can be expressed by the heat conduction equation \cite{incropera_principles_2013}

\begin{equation}
    \label{eq:heatcond}
    \rho C_p \frac{\partial T}{\partial t} = \nabla \cdot (k\nabla T) + P_{in}
\end{equation}
where $\rho$ is the mass density, $C_p$ is the specific heat capacity, $k$ is the thermal conductivity, and $P_{in}$ is the total heating power.

\section{Material Properties and Composite Design}\label{subsec:mats}
The electromagnetic modes and thermal profiles central to this work depend on the material properties of the feedstock. As such, a physical model of microwave heating requires accurate modeling of material properties. Powder melting-based technologies require careful consideration to electromagnetic properties, thermal properties, and shock resistance of all materials, with several properties being evaluated in bulk and several specific to the particles \cite{gibbons_metal_2024}. The shock resistance and melting temperatures can be considered as a threshold value for each material species separately. Due to practical limitations on FEM mesh sizes, however, properties relevant to electromagnetic mode formation and heat transfer must be treated as a bulk properties over a larger volume.

\begin{figure}[!htbp]
    \centering
    \includegraphics[width=0.75\textwidth]{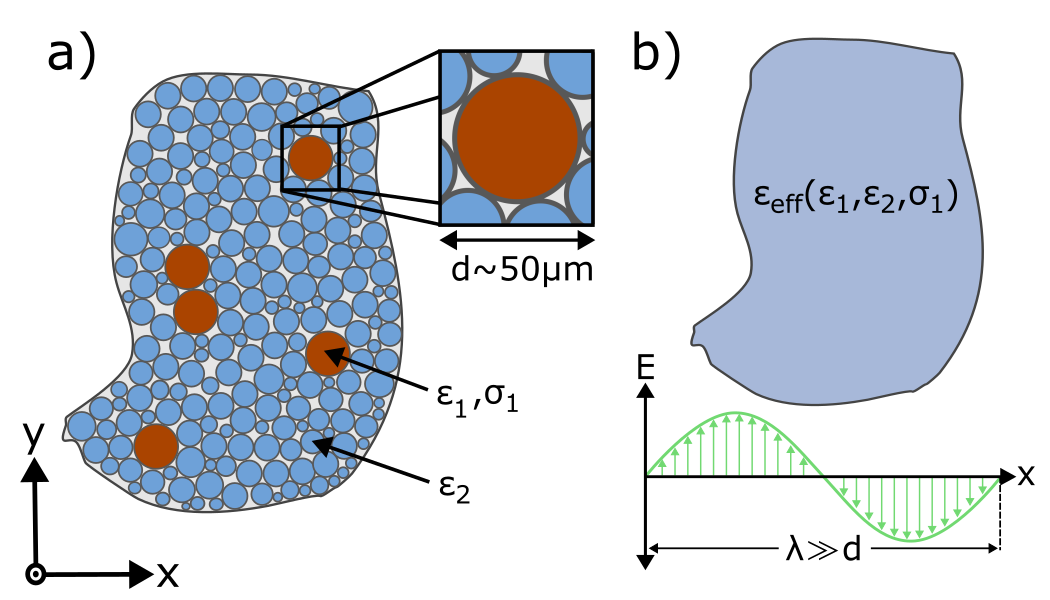}
    \caption{Diagram of the Bruggeman effective medium approximation for composite powder mixtures. a) The physical powder mixture of a ceramic powder (blue) and a metal/semiconductor (orange). Importantly, the ceramic powder is assumed to be electrically insulating. b) The uniform, effective medium seen by the microwaves when the powder dimension (d) is much smaller than the wavelength ($\lambda$).}
    \label{fig:Brug}
\end{figure}

All powder feedstocks are necessarily composites due to imperfect packing factors ($<1$), and thus require effective mean approximations to find the equivalent bulk properties of the powder volume \cite{gutierrez_electromagnetic_2019}. These approximations are valid when the radiation wavelength is much larger than the average powder grain length ($\lambda \gg d$), as scattering becomes negligible in this regime \cite{hecht_optics_2017}. In these cases, the effective mean approximation can determine the effective properties of a homogeneous powder with a vacuum medium (or inert gas) by its packing fraction or weight of the contributions of multiple powder materials in a mixture. Notably, mixing powders provides a valuable pathway to engineering the electromagnetic and thermal properties of the bulk medium, which can improve compatibility with VAM processes. In this work, both single powder feedstock and mixtures are considered. A graphical representation of the true powder material and effective bulk medium are represented in Fig.~\ref{fig:Brug}a) and Fig.~\ref{fig:Brug}b) respectively.

For a mixture of densely-packed dielectric powders we consider effective electromagnetic properties from the Bruggeman Effective Medium Approximation \cite{gutierrez_electromagnetic_2019, stroud_effective_1998} where the properties of medium 1 and medium 2 reduce to a single effective value where

\begin{equation}
    \label{eq:brug}
    f\frac{x_1 - x_{eff}}{x_1 + 2x_{eff}} + (1-f)\frac{x_2 - x_{eff}}{x_2 + 2x_{eff}} = 0
\end{equation}
for volume fraction $f$ and $x$ is any of $\varepsilon$, $\mu$, or $\sigma$. Importantly, this approximation is defined for complex values and thus naturally modifies the absorption term $\varepsilon''$ in addition to the dispersive term. This model fails, however, for dielectric-metal mixtures past the percolation threshold $f_c\approx 0.2$. In this regime, the metal grains begin to contact in extended chains and the electric absorption rises steeply. Here, the permittivity is found by \cite{bergman_critical_1977}
\begin{equation}
    \label{eq:perc_ep}
    \varepsilon_{eff} = \varepsilon_{brug} + \frac{i\sigma_{eff}}{\varepsilon_0\omega} = \varepsilon'_{eff}(1 -j\mathrm{tan}\delta_{eff})
\end{equation}
for effective permittivity $\varepsilon_{brug}$ from equation \eqref{eq:brug}, field angular frequency $\omega$, and effective conductivity \cite{bergman_critical_1977}

\begin{equation}
    \label{eq:perc_con}
    \sigma_{eff} = \sigma_m(f - f_c)^t.
\end{equation}
Here, $\sigma_m$ is the conductivity of the bulk metal and $t$ is the critical exponent which is typically assumed to be 2 for random powders. The effective thermal properties $\rho$, $C_p$, and $k$ can be approximated similarly using a modified Bruggeman Effective Mean Approximation for random, compacted powders \cite{ordonez-miranda_generalized_2010}.

This approach provides tunability of material properties, and thus allows access to regimes of higher conductivity for similar materials. A notable tradeoff for dielectric materials is the absorption and the natural reflection. Moderately conductive materials have a naturally high-absorption in small particle mixtures \cite{bohren_absorption_1998}, and thus mixing lower-loss dielectric powders with moderate conductors allows enhanced absorption. This is detailed in Fig. 2b) of the main text using several ceramics and polymers common in microwave systems, and compares their absorptive ($\mathrm{tan}\delta)$ and reflective ($|\varepsilon_r|$) properties to that of a fused-silica carbon black powder mixture. This demonstrates the value of composite powders in tuning the absorptivity to regions uncommon for non-polar dielectric materials. The alternative to poor conductors, such as carbon black, would be the addition of polar materials which would increase the absorptivity but possibly lead to unstable material properties near resonances\cite{jonscher_dielectric_1999}.

For this approach, any low-loss dielectric can be used, but the required power will be inversely proportional with the loss tangent of the material. Using these composites lowers the power requirements, and offers a pathway to designing composite solids with favourable thermomechanical.

\section{Temperature Dependence of Material Properties}\label{subsec:temp}
Electrical and thermal material properties have a strong dependence on temperature, often with notably non-linearity around phase changes \cite{yamaguchi_assignment_2011, garcia_temperature-dependent_1998}. Thermal properties of bulk ceramics and metals are often well documented, and the thermal sensitivity is included through interpolation of detailed property tables. Electromagnetic properties, however, have less reliable documentation for elevated temperatures and simplified models must be used to predict $\varepsilon'$, $\mathrm{tan}\delta$, and $\sigma$.

In non-polar materials, $\varepsilon'$ experiences only a mild temperature dependence \cite{passerat_de_silans_temperature_2009} and so we model the relationship as a linear extrapolation

\begin{equation}
    \label{eq:perm_temp}
\varepsilon'(T) = \varepsilon'(T_0) + \alpha(T - T_0)
\end{equation}
where $\alpha$ is an experimental proportionality constant and the reference temperature $T_0$ is taken to be room temperature. This linearization is usually valid for a small, experimentally verified temperature range and does not account for nonlinearities induced by phase changes. While useful in probing system sensitivities, this approximation is likely the largest source of potential non-physicality in the results. Access to detailed material data would greatly improve the reliability of the simulations, and allow direct mapping to experimental systems. Alternatively, improved theoretical models, as has been developed for materials with well defined band gaps as a function of temperature \cite{ren_novel_2025}, would also strengthen the results.

The loss tangent of temperature dependence commonly uses an activation energy model where \cite{jonscher_dielectric_1999}
\begin{equation}
    \label{eq:loss_temp}
\mathrm{tan}\delta(T) = \mathrm{tan}\delta_0 \, e^{-\frac{E_a}{k_BT}}
\end{equation}
for a high-temperature loss tangent $\mathrm{tan}\delta_0$, Boltzmann constant $k_B$, and activation energy $E_a$. The activation energy is an experimentally characterized property, known for various dielectrics including alumina \cite{vila_dielectric_1998} and silica \cite{kakiuchida_dielectric_1999}. The activation energy is the energy threshold required for lattice vibrations or defect-densities to induce absorption \cite{brailsford_anelastic_1973}. Polar materials can be accurately characterized similarly using a Debye relaxation model \cite{jonscher_new_1981}.

\begin{figure}[!htbp]
    \centering
    \includegraphics[width=0.8\textwidth]{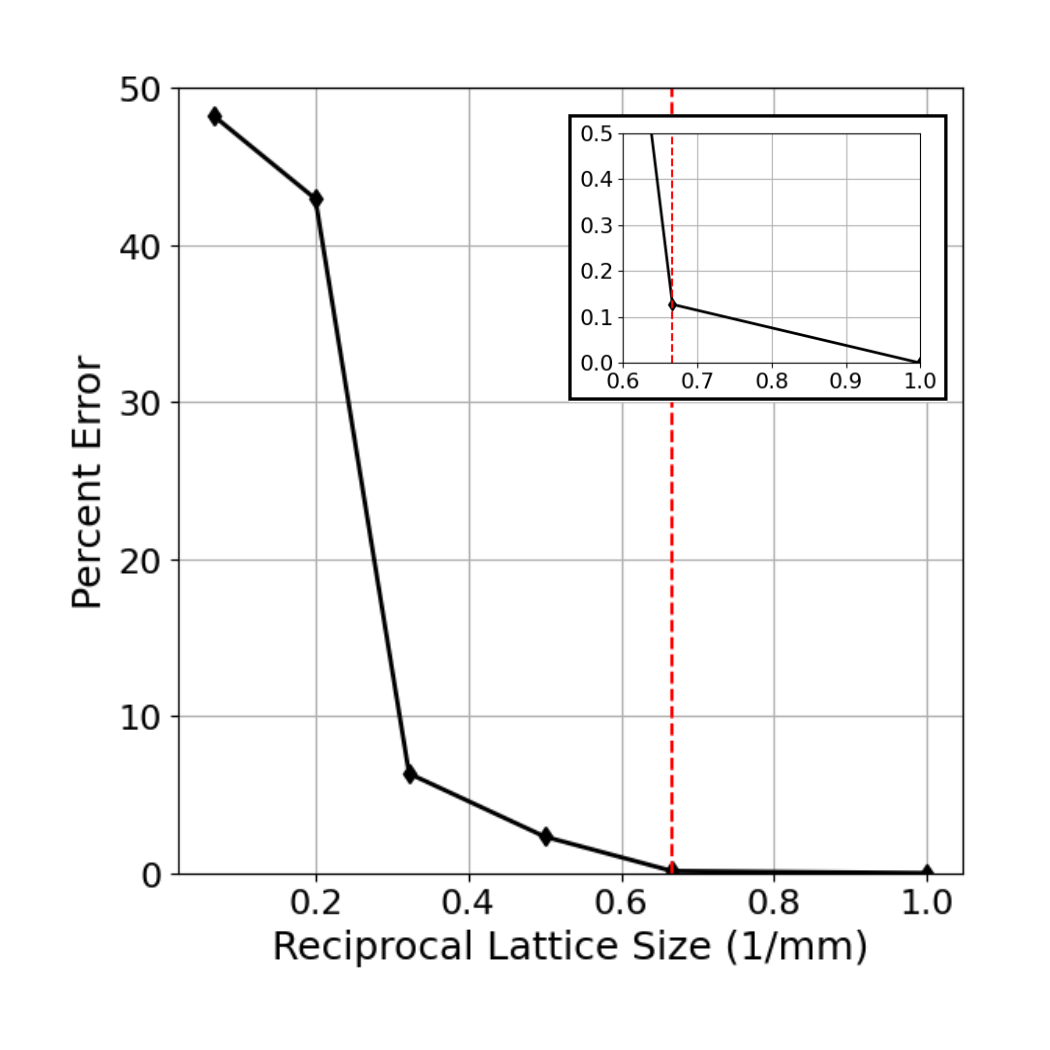}
   \caption{A convergence study of the linear resonator design. The percent error is quantified as relative error between the calculated $T_{max}$ and the $T_{max}$ acquired with the highest mesh resolution. The dotted red line shows the $\lambda/10$ condition used for meshing in all simulations. The inset shows an enhanced view of the convergent regions.}
    \label{fig:convergence}
\end{figure}

Electrical conductivity follows the same exponential relationship, called the Arrhenius model \cite{atkins_atkins_2023}, which can be used for any conductive material. These simulations only investigate copper and tungsten metals, however, and so trusted empirical formula are used, giving \cite{young_university_2012}
\begin{equation}
    \label{eq:cond_cu}
    \sigma_c = \frac{5.8e^7}{1 + (0.0039K^{-1})(T - T_0)} \, S/m
\end{equation}
for copper and \cite{young_university_2012}
\begin{equation}
    \label{eq:cond_w}
    \sigma_W = \frac{1.8e^7}{1 + (0.0045K^{-1})(T- T_0)} \, S/m
\end{equation}
for tungsten with reference temperature $T_0 = 293K$ ($0^\circ C$).

\section{Model Validation}\label{subsec:validation}
Convergence tests were used as validation of the internal consistency of the model. The presumed maximum mesh size of $\lambda/10$ was evaluated in the linear resonator design. The maximum temperature was evaluated after the last time-interval as a function of maximum mesh dimension for small timestep $t = 0.1$ms. The results are graphed in Fig.~\ref{fig:convergence}. This physical convergence is complemented by a numerical convergence predicted by the reciprocal step size decrease. Further simulations were validated by the internal COMSOL convergence plots.

\section{An Angled Interface for Field Localization in Rectangular Retro-reflectors}
\begin{figure}[!htbp]
    \centering
    \includegraphics[width=0.75\textwidth]{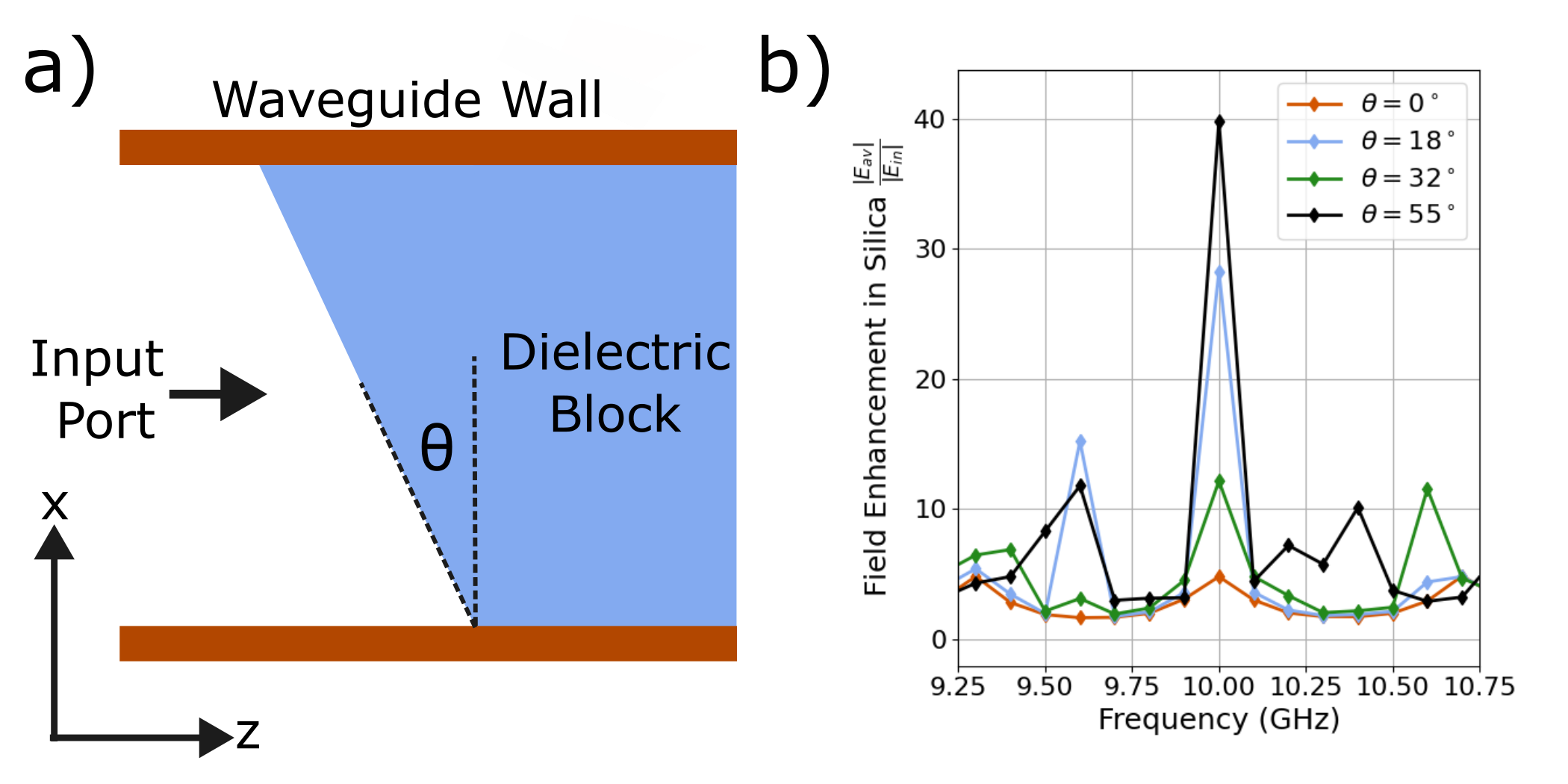}
   \caption{Optimization of electric field distribution for the boundary heater design using an angled air-dielectric interface. a) A diagram of the angled surface in the x-z plane. b) Resulting field enhancement in the silica as a function of frequency. The system was modified from the schematic in Fig. 7 in the main text, and thus the central frequency remained at 10 GHz.}
    \label{fig:angsurf}
\end{figure}

In the external printer design, an angled surface between the air/vacuum region and the silica region within the waveguide boundaries, shown in Fig.~\ref{fig:angsurf}a), breaks the paraxial symmetry and induces transverse mode mixing \cite{hill_electromagnetic_1998}. These surfaces can be used to form more uniform heating distributions and, with impedance matching, increase the energy density in the silica, reducing the power requirements for efficient heating. 

We evaluated the energy density for several interface angles and input frequencies. The results, shown in Fig.~\ref{fig:angsurf}b), highlight the increase in field concentration in the silica at the design frequency of 10 GHz, and the erratic frequency dependence of angled surfaces due to impedance matching with scattered modes. Using an angle of $55^\circ$, the power requirements are 8.3 times lower to achieve the same heating rates as with a normal interface, demonstrating one simple change which could greatly improve the experimental feasibility of the device. However, caution must be exercised when designing systems with angled reflecting surfaces. The corners of the angled surfaces can lead to mesh singularities with improper refinement, limiting physicality of the results, and the mode mixing can excite modes which concentrate fields inside the silica, heating the dielectric rather than the metal boundaries.

\bibliographystyle{unsrt}
\bibliography{paper}

\end{document}